\documentclass[aps,prL,twocolumn,longbibliography,groupedaddress,showpacs,floatfix,superscriptaddress]{revtex4-2}
\usepackage{graphicx}  
\usepackage{dcolumn}   
\usepackage[english]{babel}
\usepackage{bm}       
\usepackage{amsmath, amsthm, amssymb}
\usepackage{bbold}
\usepackage{amsmath}
\usepackage{mathrsfs}
\usepackage{array}

\usepackage{physics}
\usepackage[dvipsnames,usenames]{color}
\usepackage{soul} 
\usepackage{ulem} 

\usepackage{hyperref}
\hypersetup{
    citecolor=magenta,
    colorlinks=true,
    linkcolor=blue,
    filecolor=magenta,      
    urlcolor=blue,
    pdftitle={Overleaf Example},
    pdfpagemode=FullScreen,
    }

\emergencystretch=\maxdimen
\newcommand{\calH}{\mathcal{H}}

\def\ave#1{\langle #1\rangle}

\begin{document}

\normalem

\title{Berezinskii-Kosterlitz-Thouless Transition and Multifractal Critical Phase in Two-Dimensional Quantum Percolation}
\author{W. S. Oliveira}
\affiliation{Instituto de F\'\i sica, Universidade Federal do Rio de Janeiro Cx.P. 68.528, 21941-972 Rio de Janeiro, Rio de Janeiro, Brazil}
\author{J. Pimentel de Lima}
\affiliation{Departamento de F\'\i sica, Universidade Federal do Piau\'\i, 64049-550 Teresina, Piau\'\i, Brazil}
\author{F. A. Pinheiro}
\affiliation{Instituto de F\'\i sica, Universidade Federal do Rio de Janeiro Cx.P. 68.528, 21941-972 Rio de Janeiro, Rio de Janeiro, Brazil}
\author{R. R. dos Santos}
\affiliation{Instituto de F\'\i sica, Universidade Federal do Rio de Janeiro Cx.P. 68.528, 21941-972 Rio de Janeiro, Rio de Janeiro, Brazil}

\begin{abstract}
We present a numerical study of the two-dimensional quantum percolation model, revealing that a critical region with multifractal eigenstates mediates the transition from localized to delocalized states. 
By analyzing the mean level ratio and participation entropy, we identify two distinct transitions: a Berezinskii–Kosterlitz–Thouless (BKT) transition at the classical percolation threshold, separating the localized and critical phases, and a power-law-type transition at a larger concentration, marking the onset of full delocalization. The critical phase is characterized by multifractal eigenstates, as evidenced by the generalized fractal dimension and multifractal spectrum. Altogether, our results  establish that in the marginal two-dimensional case, the Anderson impurity model and the quantum percolation model belong to different universality classes.
\end{abstract}

\date{\today}

\maketitle


\textit{Introduction.—} The Landau-Ginzburg paradigm of phase transitions assumes the existence of an order parameter, reflecting a spontaneously broken symmetry in the ordered phase. 
This applies to finite temperature transitions, to disorder-driven purely geometric transitions (e.g.\ classical percolation \cite{stauffer18}), as well as to quantum (i.e.\ ground state) phase transitions driven by external fields, pressure, doping, disorder, etc.\,\cite{Sachdev11,Continentino17}.
This paradigm allows one to make very general statements. 
For instance, the Mermin-Wagner theorem \cite{Mermin66} rules out the existence of long-range order at finite temperatures in two dimensions for systems with continuous rotationally invariant Hamiltonians.
However, it has long been established that for a two-component order parameter, a quasi--long-range ordered (or \emph{critical}) phase can be sustained \cite{Kosterlitz73, Berezinski1971, Berezinski1972}; some examples are superconductivity~\cite{Matthew2019,Shi24,Chen25,Fontenele22}, magnetism \cite{Hu20}, and melting \cite{Halperin78,Kosterlitz16}.
The main signatures of this phase transition [usually referred to as a Berezinskii-Kosterlitz-Thouless (BKT) transition] are an exponentially diverging correlation length, $\xi\sim \exp[A/\sqrt{g-g_c}\,]$, with $A$ a constant, as the critical point, $g_c$, is approached from above, and a power law spatial decay of correlations, $\Gamma(r)\sim 1/r^{\eta(g)},\ g\leq g_c$, with a continuously varying exponent $\eta(g)$.

However, several phase transitions are not described by an order parameter, with the most notable cases being metal-insulator transitions driven by either disorder or interactions.
In this case, one cannot anticipate whether or not a critical phase may emerge, which stimulates the search in many different systems to establish a pattern. 
With this in mind, we note that within the framework of the Anderson Impurity Model (AIM) \cite{Anderson58}, tight-binding electrons experience uncorrelated random on-site energies, a type of disorder commonly referred to as \emph{diagonal} disorder. 
The two-dimensional case is marginal in the sense that the scaling theory \cite{Abrahams79} predicts that the conductivity only vanishes asymptotically at long length scales. Within the renormalization group framework this means that trajectories asymptotically flow  towards a fixed point associated with the localized phase \cite{Abrahams79}. 
While the BKT behavior of the AIM on a square lattice has been elusive, it has appeared in extensions on graphs \cite{Garcia-Mata22} and in one-dimensional many-body localization \cite{Dumitrescu19}.

A scenario similar to the AIM arises when hopping between neighboring sites is suppressed due to missing sites or interstitial defects. This \emph{off-diagonal disorder} is analogous to the classical percolation problem \cite{stauffer18}, where the system's connectivity depends on the probability, $p$, of active sites or bonds. 
Due to the connection with this purely geometrical situation, the electronic problem is referred to as the quantum percolation model (QPM). 
In this context, $p_c$ denotes the classical percolation threshold, below which there is no spanning cluster, and $p_q$ marks the quantum percolation threshold for electronic delocalization; typically, $p_c < p_q$ due to quantum interference effects~\cite{Oliveira21}.
Both the AIM and the QPM lead to localization phenomena, but the debate over the disorder threshold in the latter is still unsettled: some theoretical approaches indicate complete localization for any amount of disorder \cite{Avishai92, Soukoulis91,Shapir82,Raghavan81,Raghavan84}, while others support a finite critical disorder, above which delocalized states take over \cite{Nazareno02,Islam08,Schubert08,Gong09,Dillon14,Yu05,Oliveira25}. 

In this work, we address this long-standing controversy by presenting numerical evidence that the 2D QPM hosts a \emph{critical phase} with multifractal eigenstates in-between the localized and delocalized transitions. More significantly, we show that this intermediate phase terminates at a Berezinskii-Kosterlitz-Thouless (BKT) transition, beyond which a metallic phase emerges. Our results establish that the 2D quantum percolation model belongs to a distinct universality class from the 2D Anderson model.

\textit{Model and Method.—} Quantum percolation is usually formulated in terms of a tight-binding Hamiltonian for a single spinless electron,
\begin{align}
\label{eq:Ham}
	\calH=  - \sum_{\ave{i,j}}\left(t_{ij}a_{i}^\dagger 
a_{j}^{\phantom{\dagger}}+\mathrm{H.c.}\right),
\end{align}
in which $a_{i}^\dagger$ and $a_{j}$ are electron creation and annihilation operators respectively, $t_{ij}$ is the hopping energy and $\ave{i,j}$ denote nearest neighbors;  H.c.\ stands for hermitian conjugate. In the quantum site-percolation problem we consider a square lattice whose sites are occupied at random and independently with probability $p$, and empty with probability $q =1 - p$; thus, $t_{ij}=1$ if sites $i$ and $j$ are occupied, and $t_{ij}=0$ otherwise. 

The localized-delocalized transition may be primarily thought of as a change in the scale of the spread of the wave function over tight-binding orbitals on a lattice. 
As a result, in state space the localized regime corresponds to non-ergodicity of eigenstates, so that the spectrum does not exhibit fractal properties. By contrast, the delocalized regime is associated with ergodic eigenstates. The critical point (or critical phase, if pertinent) is expected to exhibit multifractality \cite{Kravtsov2018}.

To probe the metal-insulator transition in our system, we resort to spectral correlations that have been extensively used to characterize the transition between extended and localized phases according to their universal signatures, which depend only on the symmetry class~\cite{Evers08,luo2021universality}. 
First, Let $\epsilon_{\alpha}$ be the eigenvalues of $\calH$.
Defining the mean level ratio as by $s_{\alpha} = \epsilon_{\alpha+1} - \epsilon_{\alpha}$, we compute the \emph{ratio of adjacent gaps} as \cite{Atas13}
\begin{equation}
\label{ratio_gaps}
r_{\alpha} \;=\; \frac{\min(s_{\alpha}, s_{\alpha+1})}
                     {\max(s_{\alpha}, s_{\alpha+1})},
\end{equation}
so that averaging over $\alpha$ yields the mean level ratio $\langle r \rangle$. For delocalized systems, when time-reversal symmetry is preserved, the distribution of $r$ follows the Wigner–Dyson (WD) distribution of the Gaussian Orthogonal Ensemble (GOE) in random matrix theory, hence  $\langle r \rangle \approx 0.53$ \cite{Atas13}. 
By contrast, for a localized system, energy levels are uncorrelated so that level statistics follows the Poisson distribution, and consequently $\langle r \rangle \approx 0.39$~\cite{Atas13}. 

We characterize the structural and localization properties of the wavefunctions through the \emph{participation entropy} (PE), a Rényi-entropy which captures universal features across quantum phases~\cite{Liu25,Sierant22}. 
For a normalized state \(\ket{\Psi}=\sum_j c_j \ket{j}\) in a basis \(\{\ket{j}\}\), 
the order-\(q\) Rényi PE is defined as \cite{Luitz2014}
\begin{equation}
S_q\equiv\frac{1}{1-q}\,\ln I_q,
\label{eq:pe_renyi}
\end{equation}
with $I_q$ a generalized inverse participation ratio (IPR), 
\begin{equation}
	I_q\equiv \sum_j |c_j|^{2q};
\end{equation}	 
the usual IPR corresponds to $q=2$.
$I_q$ follows a finite-size scaling (FSS) form \cite{Ujfalusi14},
\begin{equation}
 I_q \sim \mathcal{N}^{(1-q)D_q},   
\label{eq:IPR}
\end{equation}
where $\mathcal{N}$ is the dimension of the Hilbert-space; for the present case, $\mathcal{N}=L^2$, where $L$ is the linear lattice size, and the multifractal (generalized) dimension, $D_q$, can be estimated from FSS plots.
In the localized phase, the electron can only be found on a finite number of sites, so that $I_q$ is independent of $L$; Eq.\,\eqref{eq:IPR} therefore imposes $D_q=0$ and $S_q\sim \text{const}$.
In the extended phase, $|c_j|^2\sim1/\mathcal{N}$, so that $D_q=1$ and $S_q\sim \ln \mathcal{N}$.
At the critical point (or within a critical phase) the expected behavior is multifractal, that is, $0<D_q<1$, with $D_q$ displaying a nontrivial $q$-dependence \cite{Ujfalusi14}; then, $S_q\sim D_q \ln \mathcal{N}$.
Therefore, we may take $S_q/ \ln\mathcal{N}$ as an approximant to $D_q$ even if one is in the intermendiate phases, ranging from  0 (localized) to 1 (extended).


\begin{figure}[t]
\centering
\includegraphics[width=8.6 cm]{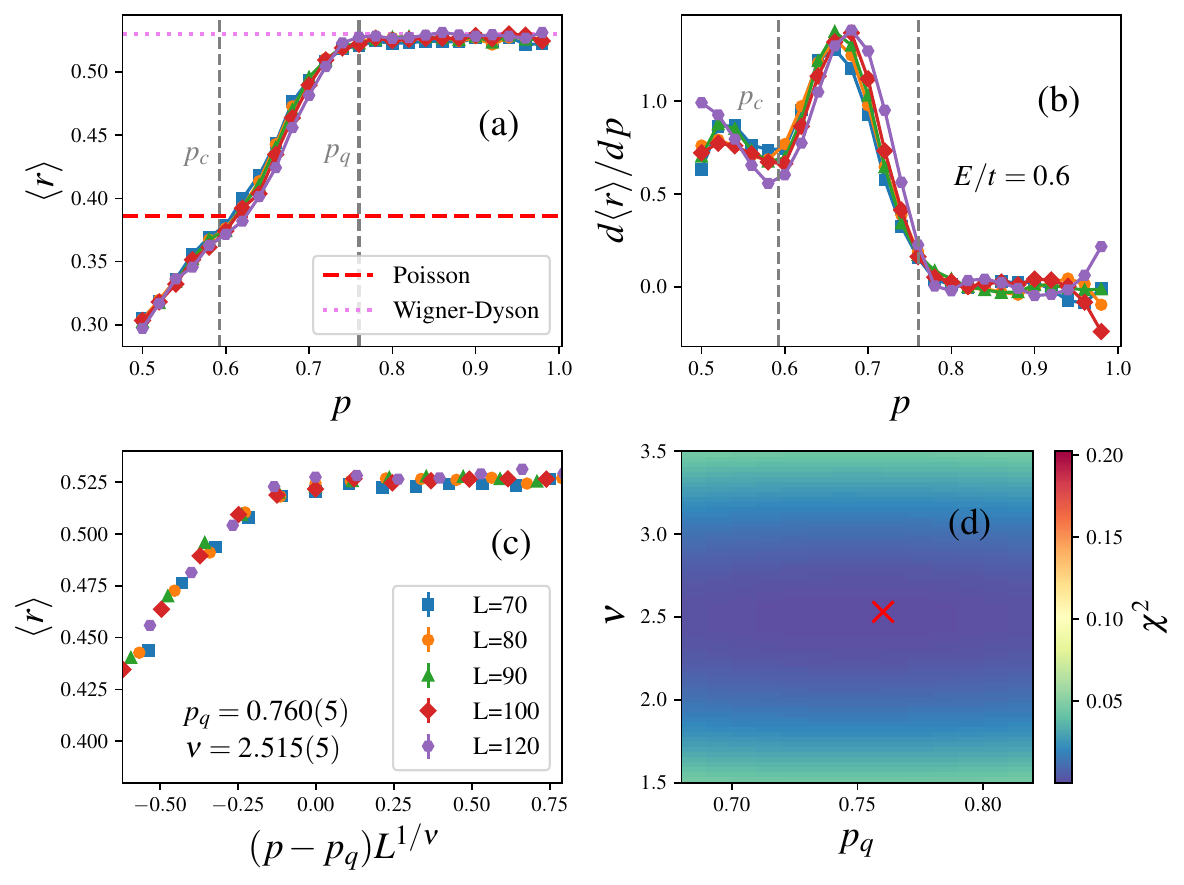}
\caption{(a) Average ratio of adjacent energy level spacings, $\langle r \rangle$, as a function of the site occupation probability $p$, for different lattice sizes $L$ 
and fixed energy interval, $E$. 
Dashed and dotted horizontal lines respectively correspond to $\ave{r}$ for Poisson and Wigner-Dyson distributions.
Dashed vertical lines indicate our estimates for the thresholds $p_c$ and $p_q$.
(b) Same as (a) but for $d\ave{r}/dp$.
(c) Finite-size scaling plot of (a).
(d) $\chi^2$ minimization to determine the optimal $p_q$ and $\nu$ (red cross) used in (c).
When not shown, error bars are smaller than data points.}
\label{fig:ratio_of_gaps}
\end{figure}

\begin{figure}[t]
\centering
\includegraphics[width=7 cm]{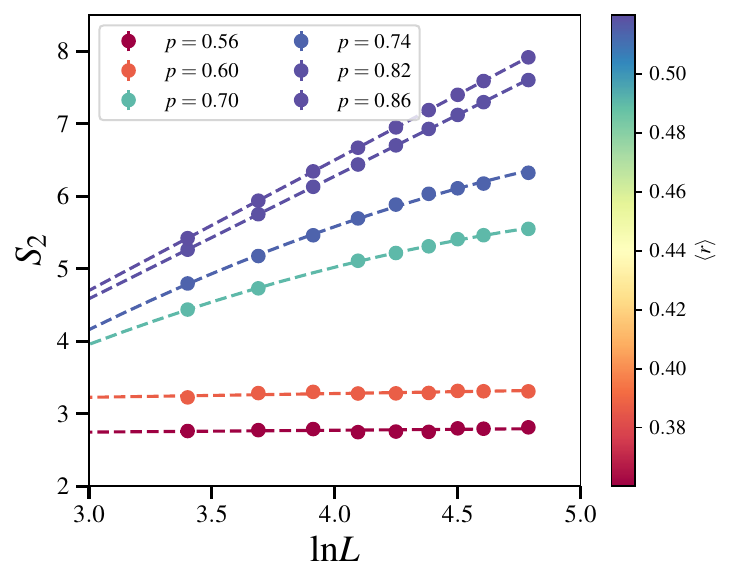}
\caption{Size dependence of the participation entropy ${S}_2$, for several occupation probabilities $p$, ranging across the different regimes appearing in Fig.\,\ref{fig:ratio_of_gaps}; see text. The colorbar encodes the mean adjacent–gap ratio $\langle r\rangle$.
Dashed lines are guides to the eye.}
\label{fig:scaling_entropy}
\end{figure}


\textit{Results.--} We start by examining the spectral statistics within the energy window $E/t=0.60\pm0.05$; other energy intervals will be considered below.
For each $L$ and $p$, we average over $10^{6}/L^{2}$ disorder realizations. Figures \ref{fig:ratio_of_gaps}(a) and \ref{fig:ratio_of_gaps}(b) display the $p$–dependence of $\langle r\rangle$ and its derivative for several lattice sizes, respectively. 
Three regimes can be clearly identified. 
At large $p$, $\langle r\rangle \approx 0.530$, consistent with a delocalized (ergodic) phase. 
As $p$ decreases, $\langle r\rangle$ drops near $p\approx 0.76$, signaling departure from ergodicity. 
Further decrease in $p$ first drives $\ave{r}$ to an inflection point near the classical percolation threshold, $p_{c}\simeq 0.5927$, and then to values below the Poisson limit, $\ave{r}=0.39$. 
This sub-Poisson dip may be attributed to the shrinking level spacings between nearly degenerate eigenstates and the ensuing level clustering, which is amplified once the lattice splits into isolated clusters below $p_{c}$ \cite{stauffer18,Iversen24}. 
A more accurate estimate of $p_q$ can be obtained with the aid of FSS analyses \cite{Fisher71,Barber83}: since $\ave{r}$ is a dimensionless variable, when plotted as a function of $L^{1/\nu}(p-p_{q})$, where $\nu$ is the localization length exponent, data for different values of $L$ and $p$ should collapse on the same curve. 
A least-squares fit provides estimates for $p_q$ and $\nu$ as illustrated in Figs.\,\ref{fig:ratio_of_gaps}(c) and \ref{fig:ratio_of_gaps}(d), which yield $p_{q}=0.760(5)$ and $\nu=2.515(5)$, in good agreement with previous findings~\cite{Oliveira25,Dillon14,Islam08,Gong09,Nazareno02,Schubert08,Yu05}.

\begin{figure}[t]
\centering
\includegraphics[width=8.6 cm]{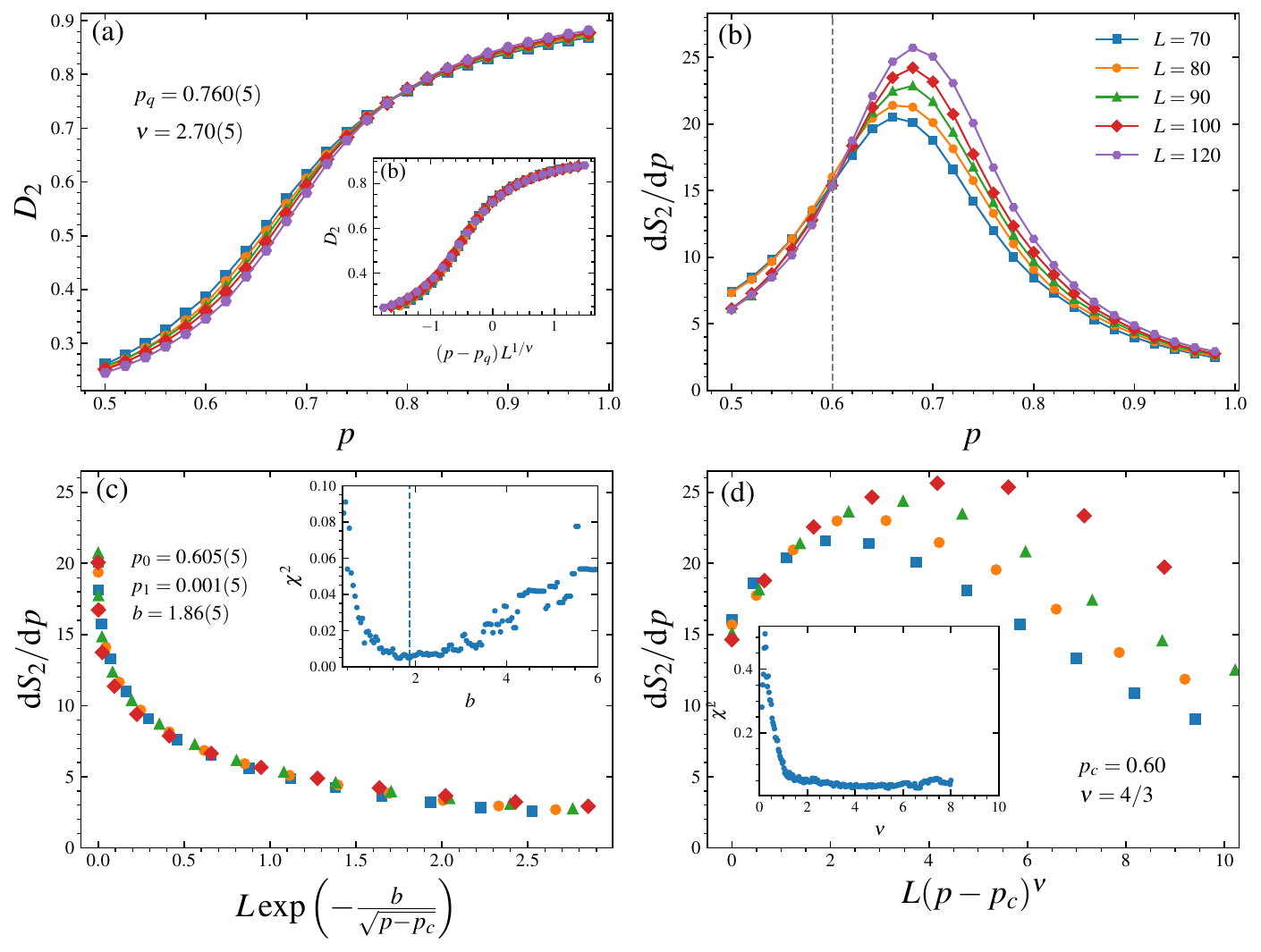}
\caption{(a) Participation entropy as a function of $p$ for different system sizes $L$. The inset shows the data collapse using a power-law correlation length. 
(b) Derivative $dS_{2}^{p}/dp$ as a function of $p$. The size-invariant crossing points are marked by the vertical line near $p_c = 0.5927$. 
(c) Data collapse of $dS_{2}/dp$ near $p_c$, assuming an exponential BKT-like localization length. The inset shows $\chi^{2}$ as a function of $b$ with the remaining fitting parameters, $p_0$ and $p_1$, fixed. 
(d) Data collapse of $dS_{2}/dp$ near $p_c$, assuming a power-law correlation length. The inset shows $\chi^{2}$ as a function of $\nu$ with fixed $p_0=0.6$ and $\nu=4/3$.}
\label{fig:part_entropy}
\end{figure}

Let us then examine the nature of the intermediate phase between $p_c$ and $p_q$ with the aid of the participation entropy, starting with $q=2$. 
Figure \ref{fig:scaling_entropy} shows the dependence of ${S}_2$ with $\ln L$, for different ranges of disorder strengths. 
Data for $0.82\leq p \leq 0.86$ show that $S_2$ approaches the ergodic limit, $S_2\sim \ln L$, or $D_2=1$, consistent with an extended phase \cite{Fabien2020}. 
In order to set a more accurate threshold for the extended phase, in Fig.\,\ref{fig:part_entropy}(a) we show the approximant $D_2$ as a function of $p$ for different system sizes, while the inset shows the resulting data collapse using the same procedure as in Fig.\,\ref{fig:ratio_of_gaps}: we get $p_{q}=0.760(5)$ and $\nu=2.65(5)$, in excellent agreement with those obtained from spectral analysis.
For $p\leq 0.6$, $S_2\sim \text{const}$, indicating a localized phase.
In the intermediate range, $0.7\leq p \leq 0.74$, the size dependence of $S_2$ is weaker than $\ln L$, indicative of multifractal behavior.
The existence of three regimes characterized by distinctive $L$ dependences is consistent with data from transmittance calculations~\cite{Dillon14}. 

The multifractal nature across a range of values of $p$ within the intermediate phase strongly suggests that this is a critical phase; this, in turn, raises the possibility that the transition between extended and critical phases belongs to the BKT universality class, which we now investigate.
From the outset, we note that Fig.\,\ref{fig:part_entropy}(a) reveals no distinctive signature of the delocalized-multifractal transition at $p_c$. 
However, a closer look at $dS_2/dp$ in Fig.\,\ref{fig:part_entropy}(b) shows that it displays size-scaling invariance exactly at $p_c$, reflecting the fact that that extended states cannot exist without a spanning cluster \cite{Odagaki80}, and at $p_c$ the support of the wave function is a geometrical fractal backbone \cite{stauffer18}.  
Let us then examine two distinct FSS variables. 
First, we consider scaling with $L/\xi_\text{BKT}$, where $\xi_\text{BKT}\sim \exp (b/\sqrt{p-p_c}\,)$, with $b$ and $p_c$ to be determined.
However, these exponential fits tend to introduce a stronger $L$ dependence in the estimates for the critical point, which are mitigated by assuming $p_c = p_0 + p_1L$ \cite{Wang2021}; thus, we end up with three parameters to be determined by a least-squares fit, namely, $b$, $p_0$, and $p_1$.
The main panel in Fig.\,\ref{fig:part_entropy}(c) shows the optimum data collapse, with the inset illustrating the minimization of $\chi^2$ with respect to $b$; note that the resulting $p_0$ lies very close to the known classical threshold $p_c\approx 0.5927$.  
Another possible FSS variable would be the usual one, i.e., assuming a power law localization length, namely $L/\xi$, with $\xi\sim (p-p_c)^{-\nu}$; as before, one seeks to minimize $\chi^2$ with respect to $p_c$ and $\nu$.
The inset in Fig.\,\ref{fig:part_entropy}(d) illustrates that for $p_c$ fixed at 0.60 $\chi^2$ displays a broad minimum, and the main panel shows that if one arbitrarily fixes $\nu=4/3$ (the exact exponent for classical percolation), the data collapse is indeed very poor; other values of $\nu$ equally fail to yield sharp minima for $\chi^2$.
The inescapable conclusion is that the transition at $p_c$ belongs to the BKT universality class, so that the whole intermediate phase is a critical one, displaying multifractality \cite{Tomasi2020}; that is, it is not a fully localized phase, hence not fully ergodic.

\begin{figure}[t]
\centering
\includegraphics[width=8.6 cm]{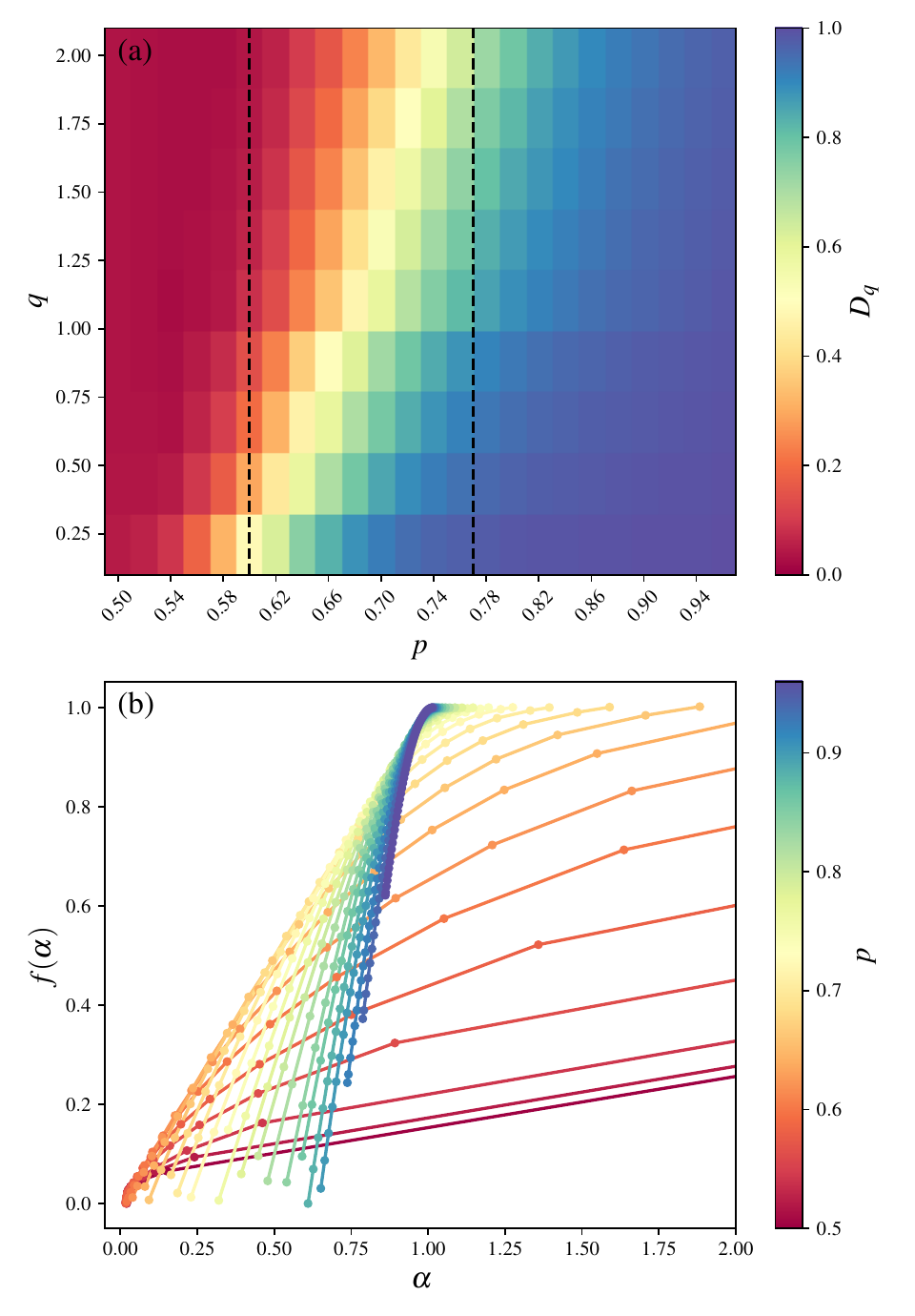}
\caption{(a) Color map of $D_q$ versus $(p,q)$. Dashed vertical lines indicate $p_c$ and $p_q$, $p_c<p_q$. 
(b) Multifractal spectra $f(\alpha)$ for 
different values of $p$, as given by the color map on the right hand side.}
\label{fig:fractal}
\end{figure}

It is instructive to follow the fractal dimension across the whole domain of $p$. Figure \ref{fig:fractal}(a) shows $D_q$ as a function of $p$ and $q$, with $0 \leq q \leq 2$. For $q<0$ the estimates tend to diverge, often lying outside the confidence intervals obtained at larger $q$, and the associated standard errors increase substantially, indicating a limited ability to distinguish the quantities.
We note that in the localized phase, $D_q$ is close to zero and nearly constant for larger $q$, consistent with wave functions confined to a finite set of sites. 
Similarly, in the delocalized phase, $D_q \approx 1$, with a weak $q$-dependence; this is the regime of spatially ergodic states. 
In the critical region, for a fixed $p$, $D_q$ decreases monotonically with $q$, the hallmark of multifractality \cite{Halsey86, Evers08}. Another fundamental and more compact measure of multifractality than $D_q$ is the multifractal spectrum, $f(\alpha)$. It captures the full content of $\tau_q=D_q(q-1)$ via a Legendre transformation. Thus, we define $\alpha=\mathrm{d}\tau_q/\mathrm{d}q$ and then evaluate $f(\alpha)=q\,\alpha-\tau_q$ \cite{Ujfalusi14}. In this representation, ergodic states yield a spectrum sharply peaked at $\alpha=1$ which converges to a delta function as $L\to\infty$; localized states show an approximately linear dependence; and genuinely multifractal states display a near-parabolic curve with maximum $f(\alpha)=1$ and the symmetry $f(1+\alpha)=f(1-\alpha)+\alpha$ \cite{Fraxanet2022}. Figure \ref{fig:fractal}(b) shows $f(\alpha)$ 
for values of $p$ across localized, critical, and fully delocalized regimes. 
We see that the shape of $f(\alpha)$ exhibits a behavior consistent with all previous predictions.

\begin{figure}[t]
\centering
\includegraphics[width=8.6 cm]{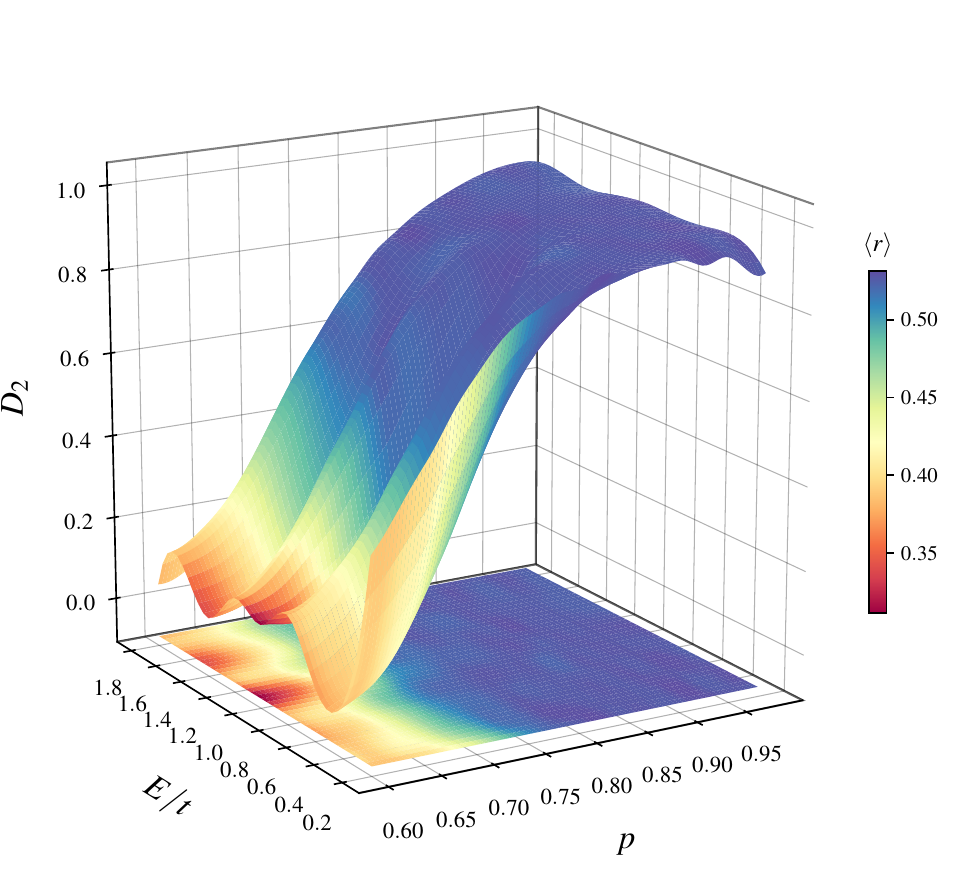}
\caption{Phase diagram in the $(p,E)$ plane. The surface shows the fractal dimension $D_2(p, E)$, while the color encodes the mean adjacent–gap ratio $\langle r\rangle$.}
\label{fig:phase_diagram}
\end{figure}
 
Finally, we recall that the results shown so far correspond to a fixed energy interval around $E/t=0.6$. 
A natural question arising is whether the results are strongly dependent on the choice of $E/t$. 
In order to address this issue, we have varied the energy interval and examined $D_2$ across the three different regimes, with the results displayed in Fig.\,\ref{fig:phase_diagram}.
We have also calculated the mean  level ratio $\ave{r}$, the values of which provide an independent check of the phases. 
We see that the qualitative behavior obtained for $E/t=0.6$ is replicated for other intervals, with the fractal dimension growing monotonically with $p$, approaching the limiting value $D_2=1$ in the extended regime at large occupation, and the localized regime below $p_c$. 
The non-ergodic and multifractal intermediate phase spans across a wide range of energy intervals, thus adding credence to the robustness of our findings. 
The slight discrepancy between $D_2$ and $\ave{r}$ noted for small and large energy intervals (an extended dominance of $\ave{r}\approx 0.4$) is attributed to the very large density of states in these spectral regions~\cite{Ujfalusi14, Oliveira25}, predominantly caused by the contribution of isolated clusters, which inflate the number of quasi-degenerate levels.


\textit{Conclusions.—} 
We have performed spectral and multifractal analyses of the two-dimensional Quantum Percolation model. 
Aided by FSS analyses, we have provided numerical evidence to settle long-standing issues relative to this problem. 
Indeed, as the concentration of active sites increases, the system enters an intermediate phase at the classical percolation threshold. 
This transition belongs to the BKT universality class, characterized by an exponentially diverging localization length as $p\to p_c^-$.
Within the intermediate phase, the participation entropy scales with a sub-linear dependence with $\ln L$, so that the system is critical, multifractal, and non-ergodic. 
The emergence of a robust intermediate phase hosting multifractal states opens a pathway toward engineering anomalous diffusion in uncorrelated disordered media \cite{ketzmerick1997determines}. 
And, finally, at the quantum percolation threshold, the wave function delocalizes. 
This transition is  characterized by a power-law diverging localization length at $p_q$, with a critical exponent $\nu=2.6\pm 0.1$ (the error stems from the two methods used). 
Thus, unlike the two-dimensional AIM, a critical amount of disorder is needed to induce localization .
Altogether, our results  establish that in the marginal two-dimensional case, the AIM and the QPM belong to different universality classes.

\section*{ACKNOWLEDGMENTS}

We thank S.L.A. de Queiroz and N.C. Costa for useful discussions.
The authors acknowledge financial support from the Brazilian agencies CAPES, CNPq, and FAPERJ.
RRdS also acknowledges grants from CNPq [314611/2023-1] and FAPERJ [E-26/210.974/2024 - SEI-260003/006389/2024]. 



\bibliography{KTPQ-final.bib}
\end{document}